\documentclass[fleqn,10pt]{wlscirep}
\usepackage[utf8]{inputenc}
\usepackage[T1]{fontenc}
\usepackage{tabularx}
\usepackage[super]{natbib}
\usepackage{booktabs}
\usepackage{ragged2e}
\usepackage{float}
\usepackage{makecell}
\usepackage{rotating}
\usepackage{newtxtext,newtxmath} % for Times New Roman font
\usepackage{array}
\usepackage{multirow}
\usepackage{geometry}
\title{Advancing a Consent-Forward Paradigm for\\ Digital Mental Health Data}
\author[1*]{Sachin R. Pendse}
\author[2]{Logan Stapleton}
\author[1]{Neha Kumar}
\author[1]{Munmun De Choudhury}
\author[2]{Stevie Chancellor}
\affil[1]{School of Interactive Computing, Georgia Institute of Technology, Atlanta, Georgia, United States, 30309}
\affil[2]{GroupLens Lab, Department of Computer Science \& Engineering, University of Minnesota - Twin Cities, Minneapolis, Minnesota, United States, 55455}
\affil[*]{sachin.r.pendse@gatech.edu}

\begin{abstract}
The field of digital mental health is advancing at a rapid pace. Passively collected data from user engagements with digital tools and services continue to contribute new insights into mental health and illness. As the field of digital mental health grows, a concerning norm has been established---digital service users are given little say over how their data is collected, shared, or used to generate revenue for private companies. Given a long history of service user exclusion from data collection practices, we propose an alternative approach that is attentive to this history: the \textbf{consent-forward paradigm}. This paradigm embeds principles of affirmative consent in the design of digital mental health tools and services, strengthening trust through designing around individual choices and needs, and proactively protecting users from unexpected harm. In this perspective, we outline practical steps to implement this paradigm, toward ensuring that people searching for care have the safest experiences possible. 
\end{abstract}
\begin{document}

\maketitle

\thispagestyle{empty}

\section*{Introduction}
The field of digital mental health is rapidly expanding. Digital modalities of mental health support have created new means for underserved populations to access care~\cite{lo2022}. They have also supported the collection of novel data via social media~\cite{chancellor2020methods}, mental health applications and wearables~\cite{abd2020effectiveness, alam2014cloud}, and online support groups~\cite{de2016discovering}. These data sources allow for insights into daily lived experiences of distress outside of clinical settings, with the potential to deeply impact the lives of those in need. Data-driven insights from past research include the prediction of symptom progression~\cite{de2016discovering}, matching individuals in distress to relevant resources~\cite{pruksachatkun2019moments, canady2021tiktok}, and identifying moments of crisis for intervention~\cite{gomes2018ethics}. In parallel, novel intervention tools are being designed by non-clinical organizations. These new tools for mental health support are being created and used at a scale and pace unparalleled by any other time in history~\cite{sinha2023understanding, wind2020covid}.

New digital approaches to support have been motivated by the potential to bring mental health research out of ivory towers and into the hands of people in distress. Diverse groups from both clinical and non-clinical domains have enthusiastically developed digital mental health tools and services, including initiatives from researchers, not-for-profit organizations, health systems, insurance companies, and technology companies. The creators of some of these tools have even applied for or received approval from the United States (U.S.) Food and Drug Administration (FDA), and are expected to get approval from the European Union’s (E.U.) Medical Device Regulation (MDR) and the United Kingdom's Medicines and Healthcare Products Regulatory Agency (MHRA)~\cite{carl2022regulating}. For many, digital mental health tools and services are a promising business opportunity, with record high investments from venture capitalists in 2020 and 2021~\cite{deangelis2023funding}.  

Digital mental health tools are fundamentally powered by data, whether through analyzing individual data to support users, identifying patterns of mental health experiences, or as a means to profit from the sale of data. Mental health data from most online tools and services is legally considered non-clinical and is thus not subject to standard health data regulations around informed use~\cite{kim2023databrokers, bossewitch2022digital}. Instead, data from digital mental health tools and services is often collected without the active and explicit consent of individuals in distress. Users have little say over how sensitive data about their mental health is collected or used, who their data is seen by, and whether their data is used to generate revenue for private actors, such as data brokers~\cite{kim2023databrokers}. Recent examples that have garnered press attention include public outcry over large language models guiding peer supporters on the Koko support platform,~\cite{nbcnews2023koko} strong negative reaction to Crisis Text Line sharing ``anonymous'' data to a for-profit spin-off company~\cite{politico2022ctl}, U.S. Senate inquiries into data privacy practices by teletherapy platforms~\cite{warren2022senate}, and U.S. Federal Trade Commission action against teletherapy companies sharing data to third party advertisers without clear consent from consumers~\cite{ftcbetterhelp2023}. These issues will only increase in scope and across domains---recent U.S. congressional hearings have raised concerns about sensitive mental health data gathered and sold by social media companies~\cite{tiktokhearings2023}.

Informed consent for data collection is a cornerstone of mental health research, in both digital and traditional formats---however, these ethical principles have not been translated to the design of many digital mental health products. Data collection from digital mental health applications is done with little oversight, a practice that is extremely different from the research protocol review that might be done by an ethical review board, such as a U.S. Institutional Review Board (IRB)~\cite{wykes2019towards}. If consent to collect data is obtained at all, it is done through opaque Terms of Service agreements or Privacy Policies~\cite{jilka2021terms, parker2019private}, which may not constitute true consent~\cite{obar2020biggest}. The issues around consent for non-clinical digital mental health data are complex, as the field is largely unregulated, and data is created as a byproduct of application or service use. These applications or services may not be mental health focused (such as social media platforms) but may contain information that can be used to infer mental health~\cite{ienca2022mental}. 

Due to a lack of regulation or standardization, individual organizations determine how they approach consent in their tools and services. Without standards around consent, organizations may rely on existing laws to decide their privacy, security, and consent practices, and in many cases, existing laws can still be misunderstood by organizations or poorly enforced by regulators~\cite{gooding2019mapping}. An unrestrained use of data could have severe consequences for the very users that care aims to protect, including discrimination in health coverage or employment, and targeted advertisements by organizations that prey on the vulnerable~\cite{bossewitch2022digital}. However, designing data collection practices around active and affirmative consent could increase the level of trust that a user has in digital mental health tools and services, and allow users to feel safer in expressing distress and seeking support---thereby improving their care and outcomes of assistance. 

In this perspective, to spur greater discussion around the role of consent and agency in digital mental health data collection, we propose an alternative approach to current practices---a consent-forward paradigm to digital mental health data. This paradigm embeds the principles of affirmative consent in the design of digital mental health tools and services, in which a digital mental health technology must ``ask for---and earn---enthusiastic approval''~\cite{im2021yes, friedman2019yes, hilgert2016burden} from a user before taking any action that involves their data. A consent-forward paradigm is attentive to the history of service users having their consent and agency ignored in data collection, may strengthen user trust through designing around individual choices and needs, and may proactively protect individuals from unexpected harm. This article outlines how new technological innovations from computer security, data privacy, and critical data studies can be leveraged to implement a consent-forward paradigm. We propose new design approaches to help realize a future in which user choice is a core part of engagements with diverse digital modalities. These approaches are rooted both in our own lived experiences engaging with mental healthcare systems, work from authors who have described their own experiences evaluating the risks of non-consensual interventions~\cite{saks2010refusing, saks2007center}, and work from authors profiling non-consensual experiences with mental healthcare systems~\cite{wipond2023your}. 

We believe that a consentful infrastructure for data collection within digital mental health spaces could allow people in distress to seek care and self-disclose freely without fear of harm, and use digital mental health technology to its fullest potential. A consent-forward paradigm could also lay the groundwork for increased trust and communication between users and practitioners, towards responsibly gaining deeper insights from data and improving care. A consent-forward paradigm is one potential approach that asks us to deeply consider the risks of technology-mediated care and how we could mitigate them. We hope that this Perspective acts as a foundation for greater discussion (and empirical investigation) of the lived experiences of service users as they engage with digital mental health tools -- and what ways that consent-forward paradigms may serve their needs. Digital mental health tools and services do not have to continue along the disempowering norms of history and present practice---rather, we argue that a different future can be realized through a \textit{consent-forward} paradigm to digital mental health data collection.

\label{sec:introduction}

\section*{Historic Perspectives on Data Consent for Mental Health}
Our consent-forward approach to digital mental health data collection is rooted in the history of how consent has been approached in mental health settings, and furthermore, legal decisions that institutionalized certain approaches to consent. This historical context highlights gaps in current approaches, as well as how those gaps may influence \textit{digital} mental health data collection practices. Historical data collection largely ignored the consent of people with mental illness, with consent often only critically considered via legal proceedings around liability or medical negligence. 

The first systematic data collection about mental health was conducted by Western asylums and the U.S. Census in the late 18th~\cite{porter1990foucault} and early 19th~\cite{andrews1998case, reid1989history} centuries. Asylum administrators and staff would collect data when an individual in distress was involuntarily committed by a family member. This process happened without patient consent, and patients needed their family's or a practitioner's advocacy to leave~\cite{hanganu2019diagnosing}. Collected data was used for organizational needs or to justify the involuntary commitment of patients (such as medical certification or warrants~\cite{hanganu2019diagnosing, wright2022asylums}) and was unrelated to patient welfare. On a national level, the U.S. Census first tracked data around mental illness in 1840. Data collectors would ask the head of each family to report the number of ``insane'' or ``idiotic'' individuals in a household (these terms are pejorative today but were the official language of the Census in 1840)~\cite{forret2016deaf}. Contemporary medical professionals noted that data collectors would overcount free Black people as insane, and supporters of slavery would use the presumed ``objectivity''~\cite{horwitz2011checkered} of systematic data collection to argue that enslavement was good for mental health~\cite{deutsch1944first}. In both data collection processes, the voices and consent of people with mental illness were excluded by design, stemming from a belief that patients did not have the mental capacity to voluntarily contribute data. 

The idea that data could be used for individual welfare was introduced with the advent of psychoanalysis---however, data was intentionally collected without consent out of the belief that it might interfere with data quality. In \textit{A Case of Hysteria},~\citealp{freud1997dora, lipton1991analyst}, Freud argued that using patient data for research that could help others in distress was a core part of a physician's duty, so long as the physician ``avoid[ed] causing direct personal injury to the single patient concerned''~\cite{freud1962fragment}, which set precedents around how case studies were analyzed and reported. However, Freud was clear in his belief that research benefits outweighed individual consent, and intentionally did not ask for consent when collecting patient data. Freud believed that patients may ``never have spoken [to him] if it had occurred to them that their admissions might possibly be put to scientific uses''~\cite{freud1962fragment}, but that the value of data to society outweighed the individual harm of not asking for consent.

Freud's emphasis on avoiding direct personal injury when collecting data also set a precedent for future legal protections against non-consensual data sharing. This process largely occurred through legal decisions around the nature of consent, as part of criminal proceedings in which clear harm was demonstrated due to data being shared non-consensually. For example, in the U.S., the legal right to the confidentiality of mental health data was first recognized through state courts in Binder v. Ruvell only in 1952~\cite{mosher2002ethical} and federally by the Supreme Court in Jaffee v. Redmond in recent times (1996)~\cite{mosher2002ethical}, both drawing on psychoanalytic theory on the value of confidentiality. Legal authorities and professional organizations later built on these cases to integrate ideas around privacy, consent, and confidentiality from psychoanalysis into proactive laws and professional codes around the world~\cite{leach1997psychological}, such as The Mental Healthcare Act 2017 (India)~\cite{bondre2021protecting} or the Philippine Mental Health Act of 2017 (the Philippines)~\cite{lally2019mental}.

These standards were further institutionalized through new privacy laws around digital health data. For example, the medical privacy rule of the U.S. Health Insurance Portability and Accountability Act (HIPAA), announced in 2000, included protections for clinical mental health data~\cite{mosher2002ethical}. Similarly, the European General Data Protection Regulation (GDPR) has special protection for data ``related to the physical or mental health of a natural person''~\cite{GDPR2016a}. The EU's GDPR protections are more wide-ranging than the clinical specificity of the US's HIPAA; however, it is unclear whether the law covers data that can be used to infer health but is not explicitly health data~\cite{ienca2022mental}. There are exceptions to the confidentiality and consent of these laws, such as the duty to warn in case of imminent harm to self or others~\cite{borecky2019reweighing}, which ties data confidentiality laws to other actions where consent can be violated to treat those experiencing crisis, such as involuntary hospitalization (also called civil commitment). Experiences with these non-consensual treatments can be deeply violating---as Saks notes: ``\textit{I know, better than most, how the law treats mental patients, the degradation of being tied to a bed against your will and force-fed medicine you didn't ask for and do not understand}''~\cite{saks2007center}.

In research spaces, the oversight of ethical review boards anchored to widely agreed upon principles (beneficence, justice, and respect for persons)~\cite{beauchamp2001principles, united1978belmont} guide researchers to ensure that future research studies do not propagate past harms and proactively consider participant consent~\cite{spellecy2021history}. In contrast, consent in data collection from mental health tools and services has often only been considered as a reaction to harm that already occurred, a consequence of court proceedings. As we describe below, a commercial, non-clinical, and mostly unregulated digital mental health landscape has propagated this reactive (rather than proactive) approach from history, and harmed users as a result. We propose an \textit{affirmative consent} paradigm for data collection in digital mental health, one that builds an infrastructure for users to have greater ownership and choice in how their data is used. Affirmative consent is the ``idea that someone must ask for---and earn---enthusiastic approval before interacting with another person''~\cite{im2021yes} and argues that consent should be voluntary, informed, revertible, specific, and unburdensome. An affirmative consent framework centers the context of a decision, providing all relevant information about the decision and its impacts on the individual. It also centers the lived experience of the person the decision affects by ensuring their experience is voluntary and not burdensome, and that decisions can be reverted. We describe each part of this framework in Table~\ref{tab:principles}, presenting how current approaches to data collection violate these principles, and the harm of those violations. 
\label{sec:background}

\section*{Data Driven Violations of Consent}
Current digital mental health tools and services leverage technological innovations to have a greater reach than past mental health services, using novel data sources to tailor services. However, these technologies can amplify non-consensual aspects of data collection in mental health services. Below, we use two contemporary case studies that demonstrate these harms---the use of algorithmic inference to support non-consensual intervention, and discriminatory targeting from monetizing digital mental health data. We draw attention to where design practices and approaches violate principles of affirmative consent (additional details in Table~\ref{tab:principles}). 

\subsection*{Algorithmically Mediated Intervention}
Online spaces, such as social media platforms or search engines, are used by individuals in distress to seek support and learn more about resources. While online platforms are not designed to provide mental health services, as a consequence of their scope, platforms are put in the position of deciding how to best connect individuals in distress to resources. Due to limited moderator resources, online social media platforms use artificial intelligence-based analyses of social media posts to identify users at risk of harm~\cite{gomes2018ethics}. In collaboration with existing services, platforms will then deploy emergency services or law enforcement~\cite{bossewitch2022digital}, paralleling the non-consensual active rescue~\cite{zeavin2021distance, bossewitch2022digital} crisis protocol undertaken by suicide hotlines. However, unlike interventions initiated by suicide hotlines, platform-based intervention can be initiated without the knowledge, let alone consent, of people in distress. Wipond describes the anecdote of an individual participating in a private Facebook support group meant for separated and divorced women, and expressing her suicidal ideation---at midnight, the police showed up at her door, with the woman ``never knowing how or why it happened''~\cite{wipond2023your}. Non-consensual active rescue is also used by digital tools and services that \textit{are} explicitly designed for mental health support--- non-consensual active rescue is initiated for users of Crisis Text Line 28 times per day on average~\cite{zeavinthirdchoice2020}.   
Non-consensual active rescue is grounded in the idea that ``anyone who is suicidal deserves aggressive intervention to keep [them] alive''~\citealp{aas2006, nsp2010}, independent of an individual's ``willingness or ability to provide consent''~\cite{nsp2010}. These protocols are enacted as a last resort and done by crisis services when all other support methods have been exhausted. Similar to laws sanctioning involuntary hospitalization, and continuing a long history of denying the agency of people with mental illness, these protocols are founded on the legal doctrine of \textit{parens patriae}. This doctrine understands an individual experiencing a mental health crisis to be unable to make decisions in their own self-interest, and the state to play a primary role in determining when intervention supersedes individual rights~\cite{testa2010civil}. 

Though potentially life-saving, non-consensual active rescue can also be terrifying and deadly. In the U.S., individuals who are seriously considering suicide are likely to be armed, which can and has been justification for shooting, traumatizing, and even killing the individual needing help~\cite{zeavinthirdchoice2020}. 23 percent of police fatalities from January 2015 to August 2020 in the U.S. resulted from police action during a mental health crisis~\cite{bossewitch2022digital}. Similar risks exist globally---from 1990 to 2016, 27\% of firearm deaths globally were from suicide~\cite{naghavi2018global, rivara2018firearm}, and suicide deaths via firearm are highly concentrated in countries in the Americas~\cite{naghavi2018global, rivara2018firearm} and Greenland~\cite{ilic2022worldwide}. Crisis intervention training for police is also limited and voluntary, with police officers expressing discomfort with their role as ``street corner psychiatrists''~\cite{teplin2000keeping} from a lack of knowledge of what might best help a person in distress. In their survey of people who have used suicide hotline services, Leppert and Jervert~\cite{leppert2022suicide} describe how respondents reported difficulties in understanding public-facing information from helplines around non-consensual active rescue and named the lack of transparency as a core reason for why they could not be honest about the extent of their suicidal ideation with hotline volunteers. 

Algorithmically mediated active rescue protocols initiated by social media platforms, health services, government agencies, and other actors~\cite{bossewitch2022digital} are similarly not transparent. One of the only available statistics on algorithmically mediated active rescue comes from a 2018 report from Meta, describing how 1000 wellness checks were initiated by the Facebook platform in 2018~\cite{bossewitch2022digital, gomes2018ethics}, but little was shared about the algorithms that triggered these checks. 
Similar to non-consensual active rescues that happen via traditional means, emergency responders still enter crisis situations with little context about what precipitated the individual's mental health crisis, and what they can do to best assist a person. Individuals in distress do not voluntarily consent to the intervention, nor are they informed about what the intervention might look like or how wide-ranging it might be. 

There is a new opportunity for algorithmic crisis prediction systems to incorporate a consentful approach to their deployment, ensuring that an individual has sufficient context for their intervention and choice for what that intervention looks like. As one supporting tool for a consent-forward approach, we describe how digital mental health researchers and practitioners could use digital psychiatric advance directives to ensure that this process centers individual consent, even when an individual is in crisis.

\subsection*{Monetization, Reidentification, and Discrimination}
Throughout history, people with mental illness had little agency, contribution, or even awareness of how their data was being used. Digital data collection continues to passively involve users, buoyed by the modern Internet's reliance on an individual's data for targeted advertising. The sale of this data can be poisonous, enabled through a lack of enforced regulatory frameworks, expansive Terms of Service policies, and unclear standards on what constitutes mental health data~\cite{ienca2022mental}. People with mental illness have little ability to voluntarily consent to their data being shared or sold, little ability to retract that consent after they have shared data, and derive little immediate benefit from the revenue generated from their data. 

This data is often not anonymized because it is not considered confidential health data, even if a user may believe otherwise~\cite{FTCComplaint2023}. This happens through the sale of data from digital mental health applications to advertisers by data brokers and can include the first and last names of individuals, their psychiatric diagnoses, the medications they take, their likelihood of mental illness, and even their contact information and home address~\cite{kim2023databrokers}. While advertisers do primarily buy this data, anyone with sufficient funds can buy this data, as demonstrated by Kim~\cite{kim2023databrokers}. Access to this data is facilitated through the opaque and expansive Terms of Service (TOS) or privacy policies found in digital mental health applications~\cite{jilka2021terms, wykes2019towards, powell2018complexity}, which allow the sharing of most data from the application. 

Users may falsely believe that their non-clinical mental health data is confidential, similar to how it is in other clinical contexts. This belief is augmented by the fact that digital mental health interfaces have been found to mislead users into thinking that all application data is confidential or not being shared, as documented in the FTC complaint against BetterHelp~\cite{FTCComplaint2023} or the FCC letter to Crisis Text Line~\cite{politico2022ctl}. Users cannot opt out of this data collection if they want to access services. Consent is thus taken with a substantial burden associated with refusing to share data, and users are not sufficiently informed about how their data is shared or about the security practices of an application. Parker at al.~\cite{parker2019private} found that nearly half of the 61 digital mental health applications they reviewed had no privacy policy whatsoever, and Iwaya et al.~\cite{iwaya2023privacy} found poor security practices among the top mental health apps in the Google Play Store.

As the FTC complaint against BetterHelp notes, deanonymization and sharing of this sensitive data can lead to impacts on an individual's ability to ``obtain and/or retain employment, housing, health insurance, or disability insurance,'' along with additional ``stigma, embarrassment, and/or emotional distress.''~\cite{FTCComplaint2023} There is the potential that this data may also be used to limit the kind of treatment that an individual can access. Data around past engagements with crisis services have been used to deny people additional care in the hospital, as part of the Serenity Integrated Mentoring (SIM) program formerly used in the UK by the NHS~\cite{kendall2023nhs}, and in trial in several U.S. states~\cite{bossewitch2022digital}. It is also possible data collection from digital mental health applications could be used for similar purposes. For example, individuals with frequent application usage may be triaged away from further accessing more involved crisis resources, especially in light of stereotypes that frequent service users are ``time-wasters''~\cite{middleton2014systematic} in clinical literature. Conversely, discrimination may happen due to a lack of data or previous use---digital mental health services may be designed such that they do not function effectively without providing user data or expansive ToS agreements. Once data is sold, users may have little recourse to reclaim their data or share in its revenue. In the following section, we describe how a consent-forward paradigm can ensure that users have full consent over how their data is used and that any shared data is anonymized and private by design. 
\label{sec:violations}

\section*{Mitigating Current Harms Through a Consent-Forward Paradigm}
As awareness around the harms of non-consensual intervention and data sharing has grown, policy initiatives have been proposed and enacted to limit user harm. Both Kim~\cite{kim2023databrokers} and Bossewitch et al.~\cite{bossewitch2022digital} argue that a comprehensive data privacy law or expansion of HIPAA could prohibit non-consensual data sharing. In April 2023, the U.S. state of Washington enacted the \textit{My Health My Data Act}, which regulates the collection, sharing, and sale of private health data, including the right to withdraw consent to data sharing~\cite{wash_state_law_1155}. Similarly, in May 2023, the FTC proposed amendments to the Health Breach Notification Rule that would include health tools and services not covered by HIPAA to notify users of data breaches~\cite{FTC_2023}. To address the potential for harm in non-consensual intervention, initiatives have been proposed that limit the use of police for wellness checks and more use of trained and unarmed mental health crisis teams~\cite{mubarak2022promoting}. We argue that these policies are sorely needed.  

The success of current and future data policy initiatives is also tied to the design of digital tools and services that implement their protections. We believe these initiatives also present an opportunity for clinicians, designers, and researchers in digital mental health to be proactive about consent within the design of tools and services. Data policy initiatives and technical design choices that protect consent could lay the groundwork for a broader infrastructure that centers user consent and agency for digital mental health data, across research and industry domains. Below, we begin by describing our vision for how principles from affirmative consent anchor a consent-forward paradigm and what an infrastructure centered on consent may look like within digital mental health data collection. We then introduce related areas of data privacy, security, and policy practices from computer science that underlie this paradigm and ensure a more consentful experience for people in distress. In each subsection, we describe how practices from these related areas could be implemented across the diverse domains in which digital mental health data collection happens. We also summarize these practices in Table~\ref{tab:recommendations}. 

We believe that the consent-forward paradigm we present provides one potential approach to more deeply consider user agency and current risks when designing digital applications and data collection tools. However, our recommendations should not be taken as a blanket prescription for any case of mental health data collection. Consent is highly contextual, and other approaches may be more amenable to a user (and are crucial to further study). Our perspective is highly tied to our own lived experiences and research paradigms, and might be very different from those of a population with identities and mental health needs that are very different from our own. We propose the following paradigm and associated practices as a beginning point for practitioners and researchers to more deeply consider issues around consent and agency in their design processes and implementations.  

\subsection*{Infrastructuring Consent: Trustworthy Data Collection Landscapes}
It is clear that data from digital mental health applications and services is valuable to understand lived experiences with mental illness and identify where services may not meet user needs. For example, first looking to healthcare applications, sensor data from wearable health applications can point to precursors to suicide risk~\cite{alam2014cloud}. Looking to research domains, demographic information may be used to understand how underrepresented groups use platforms differently~\cite{rochford2023leveraging}. Looking to commercial applications, engagement data may be used to understand whether tools and services create sustained and long-term value~\cite{borghouts2021barriers} for users. Current norms around data collection from these services are based on opt-in or opt-out models of consent, in which users are prompted to decide whether they want to share (opt-in) or not share (opt-out) their data when they begin to use a service. User preferences around data collection are often stored long-term, even with the EU's GDPR data protections in place~\cite{sanchez2019can}, and it has been debated whether this one-time interaction this truly constitutes consent~\cite{obar2020biggest}. Opt-in and opt-out models of consent are also widely used in medical settings---for example, the UK's national data opt-out policy allows ``[patients] to choose if they do not want their confidential patient information to be used for purposes beyond their individual care and treatment''~\cite{NHS_Digital_2023}.
Similarly, experts debate whether opt-in and opt-out models constitute actual consent in health settings, such as in the case of organ donation~\cite{mackay2015opt}.

Changing a landscape in which digital mental health data is largely collected and sold without user consent requires changes in technology design, policy reform, and norms around how users and their data are understood across diverse domains---what we call changing the \textit{infrastructure} of consent. A consent-forward approach rooted in affirmative consent is one path to ensure that data is collected with the active participation of users in contributing their data. As we present in Table~\ref{tab:principles}, a consent-forward approach is one in which digital service users have five criteria: users can \textbf{voluntarily} provide their consent to sharing data, are fully \textbf{informed} about how their data will be used before it is shared, can \textbf{reverse} their decision to share their data, can choose what \textbf{specific} types of data they want to share, and are not required to agree to a \textbf{burdensome} terms of service that requires data sharing to access services. Creating a landscape of digital mental health tools and services that accommodates these principles will require policy interventions~\cite{kim2023databrokers}, empirical research to understand what forms of requests for consent are not so burdensome that they dissuade people from accessing care, and innovative technology design to ensure that users can manage how their data is being used, should they choose to use a service. We understand these design, norms, and policy shifts around consent in digital mental health to broadly lay the \textit{infrastructure} for a consent-forward approach.

We believe that a core part of transitioning to a consent-forward approach is greater communication between digital mental health service providers (including both research organizations and commercial entities) and potential service users. This could naturally happen individually, with users able to quickly communicate their preferences for how their data is collected and used. However, it is also important that this happens on an institutional level, through lived experience advisory boards. Several mental health service providers and organizations (such as the U.S. 988 Suicide and Crisis Lifeline~\cite{988lifeline_lived_experience} and the U.S. Substance Use and Mental Health Services Administration~\cite{samhsa_advisory_councils}) have lived experience advisory boards that they consult to ensure that service provision is acceptable to diverse service users. 

In transitioning to a consent-forward approach, it is possible that underrepresented groups may feel uncomfortable voluntarily sharing their data because of a lack of trust in researchers, a pattern observed in participation in research studies~\cite{george2014systematic}. Similarly, in commercial domains, users may not trust private corporations to safeguard their data, particularly given data breaches and violations of consent in the past~\cite{FTCComplaint2023, politico2022ctl}, as well as the profit associated with the sale of sensitive data. However, we understand a consent-forward approach to be an opportunity for the creators of digital mental health tools and services to build stronger trust among potential digital service users in light of past harms and violations, such that users feel safe and comfortable voluntarily sharing their data. As George et al.~\cite{george2014systematic} note, one method to mitigate the underrepresentation of minority individuals in research is through consent being treated as ``an ongoing process—a dialogue—rather than a discrete act of choice that takes place in a singular moment in time.'' A similar dialogue-based approach may be successful in enabling data collection from digital mental health tools and services without compromising user safety or experience. 

Another practice that could help build trust at an institutional level is institutional support for the formation of patient advocacy groups and lived experience advisory boards. Lived experience advisory boards to facilitate dialogue, helping creators of tools and services understand where repeated requests for consent may be burdensome and not worth the risk of dissuading people from seeking support. Though a lived experience panel is out of scope for this Perspective piece, an advisory board or panel could assist in empirically validating what modifications are needed to our approach to be successful and encourage people to feel safer accessing support, and ensure that data is responsibly learned from (in line with member checking approaches in qualitative research~\cite{mckim2023meaningful, lincoln1985naturalistic}). Along with the mechanisms we describe below (and outlined in Table~\ref{tab:recommendations}, an affirmative consent approach that treats consent to data contribution as a dialogue between service providers and users could allow for safer access to digital mental health support services.

\subsection*{Encoding Consent: Digital Psychiatric Advance Directives (DPADs)}
Individuals may engage with care under severe distress with impaired decision-making capabilities~\cite{sastre2021decision}. Users in severe distress may also engage with multiple forms of care online, including online crisis chat services, social media-based resources, or search engines. We propose \textit{digital psychiatric advance directives} (DPADs) as one way that users have full consent over their engagements with online tools and services, including what data is collected from those engagements. Traditional psychiatric advance directives are legal documents that ``[allow] a person with mental illness to state their preferences for treatment before a crisis.'' Along similar lines, DPADs could encode consent into the design of digital tools and services. A DPAD could contain information similar to a traditional psychiatric advance directive, including close contact information, designated legal proxies, and exact information about treatment plans and their execution. Similar to mobile Medical IDs,~\cite{sandler2020mobile} DPADs could be stored privately on mobile phones, with functionality enabling directives to be accessed by health providers when an individual indicates that they are in crisis or are incapacitated. 

DPADs also allow users to determine what data they are comfortable sharing before engaging with digital tools, services, and commercial applications. We envision that applications may ask an individual for their DPAD before engagement and collect data as per the directive---if a directive is not provided, a consent-forward paradigm would suggest that data not be collected. DPADs thus fulfill the principles of affirmative consent by allowing an individual to \textbf{voluntarily} choose the \textbf{specific} forms of data they are comfortable sharing and do so while not in a \textbf{burdensome} context (such as being in crisis) and while \textbf{informed} of how data might be used. 

Secure methods could be used to implement DPADs, ensuring that individuals can revoke or \textbf{revert} their previously provided consent around data collection. Smart contracts are one such example, consisting of digital contracts that enable the contract creator to share private information with any individual or organization based on the terms of the contract~\cite{gupta2021securing, yue2016healthcare} without intervention from a third party. Smart contracts could be used to ensure that later changes in preferences around data collection in a DPAD are automatically reflected in data collected from a user's past engagements with care. We envision that DPADs would thus take on a standardized and interoperable form across services, but with additional service-by-service questions relevant to new and future digital mental health support forms. However, empirical research is needed to understand what a base standardized DPAD should include, based on what kinds of data could be collected and user perspectives on collecting that data.

To accommodate a widespread use of interoperable DPADs, digital mental health services would have to design their data collection and analysis techniques to be sensitive to the directives of users, in line with changes made to comply with the ``right to erasure''~\cite{sarkar2018towards} in the GDPR. Doing so would encode consent into the design of digital tools and services, ensuring that users fully consent to their engagements with care. While the integration of DPADs into digital mental health tools and services might be complex, there is a strong ethical and commercial incentive to leverage them---users may feel more safe expressing distress freely when using mental health applications, as they would not have to fear action being taken without their awareness or consent. 

\subsection*{Fortifying Consent: Security Practices for Large Scale Analyses}
One motivation for collecting mental health data is the potential to identify patterns of distress that traditional research methods may not catch or to better understand whether digital interventions are working well to ease user distress. However, data collection comes with risks. Data can be collected from commercial applications without sufficient anonymization~\cite{kim2023databrokers} and can be integrated with other publicly available data sources (such as social media data) to reidentify users~\cite{simon2019assessing}, even if an application may assure a user that data is deidentified. DPADs may allow users to delineate what engagement data they are comfortable sharing. Additionally, ensuring that users trust the data protection processes of tools and services could make them more willing to contribute their data, even if they have poor experiences. To protect users, we describe three approaches from the field of usable security and privacy that ensure that data can be collected and analyzed while minimizing risk. The field of usable security and privacy investigates these human-focused (or ``usable'') practices in trust, security, and safety to make interactions more secure, private, and transparent. The efficacy of each of the data security approaches we describe below can also feasibly be audited by independent organizations, further strengthening trust in services across research and industry domains. 

\textit{Differential privacy.} Differential privacy is an algorithmic approach to anonymization that ensures that overall patterns that describe the data stay consistent, while individual data points do not yield sufficient information for re-identification. This is done by adding noise or altering feature values so predictions are still accurate, but individual datapoints are not. Differential privacy approaches are effective at anonymizing health data broadly~\cite{niinimaki2019representation, liu2020blockchain, lv2021security} and digital mental health data specifically~\cite{roy2022interpretability, basu2021benchmarking}. 

\textit{Federated learning.} Digital mental health tools and services often collect data from centralized or third-party servers~\cite{iwaya2023privacy}. This centralization of data and poor security practices risk privacy violations, particularly given the sensitivity of mental health data. With federated learning, predictive models could be trained locally on user devices---instead of sharing user data, model weights from users are sent to a central hub for cross-user analysis. A federated learning approach has been found to be helpful in healthcare broadly~\cite{antunes2022federated, brauneck2023federated}, and in digital mental health applications~\cite{basu2021benchmarking}.

\textit{End-to-end encryption.} Confidential communication is of the utmost importance in mental healthcare. However, many mental health applications (including some facilitating access to support groups) do not encrypt user communications~\cite{iwaya2023privacy}. %, leaving users vulnerable to data breaches or re-identification. 
End-to-end encryption ensures that the platform owners cannot read or make inferences from messages, protecting sensitive data from being exposed in breaches---strictly senders and recipients can access messages locally. It is important that the meaning of encryption be clear to users via application interfaces, as users may not have an understanding of how end-to-end encryption works (even if they are told that a platform is end-to-end encrypted)~\cite{dechand2019encryption}.

Together, these three approaches guard affirmatively consentful approaches by ensuring that users can trust that the most sensitive parts of their data cannot be shared with others by design.

\subsection*{Systematizing Consent: Data Governance Strategies}
Research on the willingness of individuals to share their health data for scientific research has demonstrated widespread public support for data sharing~\cite{kalkman2022patients}. However, as Kalkman et al.~\cite{kalkman2022patients} note, this support can be dependent on whether there exists transparency around how data is being used by researchers, privacy and security protections, and some ability for contributors to govern how their data is used. Looking specifically to digital mental health contexts, recent work~\cite{sieberts2023young, bossewitch2022digital} has also emphasized the importance of participatory data governance in the voluntary contribution of mental health research data. For example, Sieberts et al.~\cite{sieberts2023young} found that young users of digital mental health applications in India, South Africa, and the United Kingdom are enthusiastic about contributing their data to research, so long as they can govern why and how that data is being used. The value these service users associate with data governance is a counterpoint to historical patterns of exclusion from mental health data collection practices.

Historically, users have had little ability to consent to their data being used to support policy initiatives. A consent-forward paradigm encompasses not only an individual's ability to consent to specific parts of their data being shared, but the ability of users to \textit{collectively} consent to how their data is used. This is a core role that data governance, or determining ``who has authority and control over data and how that data may be used''~\citealp{olavsrud2021data, ada2021participatory}, plays in a consent-forward paradigm. We argue that mental health data governance is crucial to ensuring that data use immediately benefits people in need and that benefits between users and digital mental health companies are symmetric. Core parts of governance in historical service user movements included ``equal power of members, horizontal decision-making, democratic structure'' and ``restoring individual and collective power''~\cite{campbell2005historical}. Mental health data governance can be designed to support these values, such as how peer supporters could use computational tools (such as Decentralized Autonomous Organizations) for consensus-based decision-making around code of conduct~\cite{thrul2022web3}. Peer support groups could function similarly to data cooperatives~\cite{temanararaunga}, using similar tools around democratic decision-making to collectively determine whether their data will be shared, to whom, and the boundaries to that data sharing. Feygin et al.~\cite{feygin2021data} have proposed that data dividends could be paid to people who voluntarily contribute their data. A similar approach could be used to compensate user groups who do decide to contribute their data, to academic or industry researchers. 

We provide these examples of potential data governance structures as starting points that demonstrate it is valuable to have people with lived experience govern how their data is used, for users, designers, and practitioners. However, the most optimal structure is highly dependent on the values of the users and groups contributing data. Integrating these considerations of mental health data governance into the design of digital tools and services ensures that users can consent to not only the data they share individually, but also how their data is used, and by whom, without them suffering a lack of service provision or resources.  
\label{sec:consent_forward_approach}

\section*{Summary and Synthesis}
\label{sec:summary}
In this perspective, we describe how a consent-forward paradigm to data collection in digital mental health can be one starting point to ensuring that users have greater choice and are protected from harm as they engage with digital tools and services. Digital mental health tools and services are increasing in popularity around the world, and though our perspective primarily discusses Western legal frameworks, we believe that aspects of a consent-forward paradigm to data collection can be adapted to bring value to diverse geographical and cultural contexts. A consent-forward paradigm is responsive to service user marginalization in past data collection practices, and endeavors to incorporate consent into every part of data collection, analysis, and governance. There will continue to be tradeoffs between the risks of digital mental health data collection, and the benefits provided by analyses of collected data. However, through a greater integration of consent-forward practices, we believe that users in distress can feel more comfortable engaging with digital mental health tools and services, and have the safest experiences possible when searching for care.

\section*{Acknowledgements}
Part of this work was supported by funding from the National Institutes of Mental Health (R01MH117172 and P50MH115838), from the National Science Foundation (1952085 and 2000782), from an unrestricted gift from Google, from the American Foundation for Suicide Prevention, and from the Microsoft AI for Accessibility Initiative. Any opinions, findings, and conclusions or recommendations expressed in this material are those of the authors and do not necessarily reflect the views of Google. 

\section*{Contributions}
The authors contributed in the following manner in the preparation of the manuscript: \textit{Concept and design:} Pendse, Stapleton, Kumar, De Choudhury, Chancellor. \textit{Drafting of the manuscript:} Pendse, Stapleton, Kumar, De Choudhury, Chancellor. \textit{Critical revision of the manuscript for important intellectual content:} Pendse, Stapleton, Kumar, De Choudhury, Chancellor.

\section*{Competing Interests}
The authors declare no competing interests. 

\begin{table}[!ht]
    \centering
    \caption{The affirmative consent framework, with specific examples of where violations of the framework occur in algorithmically-mediated interventions and monetization of data}
    \renewcommand{\arraystretch}{1.2}
    \begin{tabularx}{\textwidth}{p{2cm}XXX}
        \toprule
        \textbf{Principle of Affirmative Consent } & \textbf{Description} & \textbf{Case Study 1: Algorithmically Mediated Interventions} & \textbf{Case Study 2: Impacts of Monetization} \\
        \midrule
        \textbf{Voluntary} & Digital service users provide consent freely and enthusiastically & Users in distress do not choose to have their data analyzed for signs of crisis, and may not even know such analyses are occurring.  & Users may not be aware that data from their engagements is being collected, and may even be misled to believe their engagements are confidential by digital interfaces.   \\
        \midrule
        \textbf{Informed} & Digital service users are informed about the context of their decision before providing consent  & Users are provided little information or transparency around what happens after non-consensual active rescue protocols are engaged. & Many digital tools and services do not have privacy policies or ToS agreements---if they do, these are too opaque and difficult for an average user to understand. \\
        \midrule
        \textbf{Revertible} & Digital service users can revoke consent at any time & Non-consensual active rescue protocols do not ask for consent, nor do they give opportunity for it to be revoked. & Once data is shared, there is little opportunity for users to unshare the data, as ToS agreements can often be signed as part of one-time engagements (such as with crisis services).   \\
        \midrule
        \textbf{Specific} & Digital service users consent to specific actions, rather than a series of actions & Per the doctrine of \textit{parens patriae}, users are not understood to be able to make specific decisions around the kind of care they want, and may be involuntarily hospitalized. & Users have little ability to choose exactly what data they want to share (or not share).  \\
        \midrule
        \textbf{Unburdensome} & Digital service users should not feel a burden to provide consent when they would rather say no & Users often engage with digital tools and services in a state of extreme and burdensome distress, which can affect their perception and decision-making processes.  & Users sign ToS agreements or review privacy policies while in a state of extreme distress, and service provision may be contingent on their willingness to contribute data.  \\
        \bottomrule
    \end{tabularx}
    \label{tab:principles}
\end{table}
\newcolumntype{Y}{>{\centering\arraybackslash}X}
\begin{table}[!ht]
    \centering
    \caption{Data privacy, security, and policy practices for consentful engagements with digital tools and services}
    \begin{tabularx}{\textwidth}{@{} >{\hsize=.5\hsize}X >{\hsize=1.5\hsize}X @{}}
    \toprule
        \textbf{Data Privacy, Security, and Policy Practices} & \textbf{Examples of Usage} \\
    \midrule
        \textbf{Digital Psychiatric Advance Directives (DPADs)} & DPADs are documents that describe an individual's preferences around their treatment if they are in crisis. They include clear guidelines around what data an individual feels comfortable sharing, as decided by the individual prior to any experience of crisis or severe distress. DPADs can be utilized by digital tools and services to ensure that any subsequent interventions or data collection center the consent and choice of the user. \\
    \addlinespace
        \textbf{Differential Privacy} & Differential privacy is an algorithmic approach to anonymization that ensures that individual data points are difficult to re-identify, while broader trends in the data stay constant. Differential privacy techniques can be used to ensure that data cannot be easily reidentified, even in case of data breaches. \\
    \addlinespace
        \textbf{Federated Learning} & Federated learning is an approach to machine learning in which predictive analyses are done at the user level, with model weights being used for large-scale analyses (instead of user data). This approach ensures that data used for machine learning analyses is similarly resilient to data breaches. \\
    \addlinespace
        \textbf{End-to-End Encryption} & End-to-end encryption is an approach to securing communications that ensures that platform owners cannot read or make inferences from messages, and that strictly senders and recipients can access messages locally. The use of end-to-end encryption secures private and sensitive communications between users, limiting them from potential harm. \\
    \addlinespace
        \textbf{Data Governance} & A consideration of data governance, or who has authority over how and why data is used, can ensure that data is used towards the benefit of users. Data governance strategies can be based on past service user movements, including democratic methods of decision-making and centering individual and collective power. \\
    \bottomrule
    \end{tabularx}
    \label{tab:recommendations}
\end{table}


\begin{thebibliography}{100}
\urlstyle{rm}
\expandafter\ifx\csname url\endcsname\relax
  \def\url#1{\texttt{#1}}\fi
\expandafter\ifx\csname urlprefix\endcsname\relax\def\urlprefix{URL }\fi
\expandafter\ifx\csname doiprefix\endcsname\relax\def\doiprefix{DOI: }\fi
\providecommand{\bibinfo}[2]{#2}
\providecommand{\eprint}[2][]{\url{#2}}

\bibitem{lo2022}
\bibinfo{author}{Lo, J.} \emph{et~al.}
\newblock \bibinfo{title}{Telehealth has played an outsized role meeting mental health needs during the covid-19 pandemic}.
\newblock \bibinfo{howpublished}{KFF} (\bibinfo{year}{2022}).

\bibitem{chancellor2020methods}
\bibinfo{author}{Chancellor, S.} \& \bibinfo{author}{De~Choudhury, M.}
\newblock \bibinfo{journal}{\bibinfo{title}{Methods in predictive techniques for mental health status on social media: a critical review}}.
\newblock {\emph{\JournalTitle{NPJ digital medicine}}} \textbf{\bibinfo{volume}{3}}, \bibinfo{pages}{43} (\bibinfo{year}{2020}).

\bibitem{abd2020effectiveness}
\bibinfo{author}{Abd-Alrazaq, A.~A.}, \bibinfo{author}{Rababeh, A.}, \bibinfo{author}{Alajlani, M.}, \bibinfo{author}{Bewick, B.~M.} \& \bibinfo{author}{Househ, M.}
\newblock \bibinfo{journal}{\bibinfo{title}{Effectiveness and safety of using chatbots to improve mental health: systematic review and meta-analysis}}.
\newblock {\emph{\JournalTitle{Journal of medical Internet research}}} \textbf{\bibinfo{volume}{22}}, \bibinfo{pages}{e16021} (\bibinfo{year}{2020}).

\bibitem{alam2014cloud}
\bibinfo{author}{Alam, M. G.~R.}, \bibinfo{author}{Cho, E.~J.}, \bibinfo{author}{Huh, E.-N.} \& \bibinfo{author}{Hong, C.~S.}
\newblock \bibinfo{title}{Cloud based mental state monitoring system for suicide risk reconnaissance using wearable bio-sensors}.
\newblock In \emph{\bibinfo{booktitle}{Proceedings of the 8th International Conference on Ubiquitous Information Management and Communication}}, \bibinfo{pages}{1--6} (\bibinfo{year}{2014}).

\bibitem{de2016discovering}
\bibinfo{author}{De~Choudhury, M.}, \bibinfo{author}{Kiciman, E.}, \bibinfo{author}{Dredze, M.}, \bibinfo{author}{Coppersmith, G.} \& \bibinfo{author}{Kumar, M.}
\newblock \bibinfo{title}{Discovering shifts to suicidal ideation from mental health content in social media}.
\newblock In \emph{\bibinfo{booktitle}{Proceedings of the 2016 CHI conference on human factors in computing systems}}, \bibinfo{pages}{2098--2110} (\bibinfo{year}{2016}).

\bibitem{pruksachatkun2019moments}
\bibinfo{author}{Pruksachatkun, Y.}, \bibinfo{author}{Pendse, S.~R.} \& \bibinfo{author}{Sharma, A.}
\newblock \bibinfo{title}{Moments of change: Analyzing peer-based cognitive support in online mental health forums}.
\newblock In \emph{\bibinfo{booktitle}{Proceedings of the 2019 CHI conference on human factors in computing systems}}, \bibinfo{pages}{1--13} (\bibinfo{year}{2019}).

\bibitem{canady2021tiktok}
\bibinfo{author}{Canady, V.~A.}
\newblock \bibinfo{journal}{\bibinfo{title}{Tiktok launches mh guide on social media impact on teens}}.
\newblock {\emph{\JournalTitle{Mental Health Weekly}}} \textbf{\bibinfo{volume}{31}}, \bibinfo{pages}{5--6} (\bibinfo{year}{2021}).

\bibitem{gomes2018ethics}
\bibinfo{author}{Gomes~de Andrade, N.~N.}, \bibinfo{author}{Pawson, D.}, \bibinfo{author}{Muriello, D.}, \bibinfo{author}{Donahue, L.} \& \bibinfo{author}{Guadagno, J.}
\newblock \bibinfo{journal}{\bibinfo{title}{Ethics and artificial intelligence: suicide prevention on facebook}}.
\newblock {\emph{\JournalTitle{Philosophy \& Technology}}} \textbf{\bibinfo{volume}{31}}, \bibinfo{pages}{669--684} (\bibinfo{year}{2018}).

\bibitem{sinha2023understanding}
\bibinfo{author}{Sinha, C.}, \bibinfo{author}{Meheli, S.}, \bibinfo{author}{Kadaba, M.} \emph{et~al.}
\newblock \bibinfo{journal}{\bibinfo{title}{Understanding digital mental health needs and usage with an artificial intelligence--led mental health app (wysa) during the covid-19 pandemic: Retrospective analysis}}.
\newblock {\emph{\JournalTitle{JMIR Formative Research}}} \textbf{\bibinfo{volume}{7}}, \bibinfo{pages}{e41913} (\bibinfo{year}{2023}).

\bibitem{wind2020covid}
\bibinfo{author}{Wind, T.~R.}, \bibinfo{author}{Rijkeboer, M.}, \bibinfo{author}{Andersson, G.} \& \bibinfo{author}{Riper, H.}
\newblock \bibinfo{journal}{\bibinfo{title}{The covid-19 pandemic: The ‘black swan’for mental health care and a turning point for e-health}}.
\newblock {\emph{\JournalTitle{Internet interventions}}} \textbf{\bibinfo{volume}{20}} (\bibinfo{year}{2020}).

\bibitem{carl2022regulating}
\bibinfo{author}{Carl, J.~R.} \emph{et~al.}
\newblock \bibinfo{journal}{\bibinfo{title}{Regulating digital therapeutics for mental health: Opportunities, challenges, and the essential role of psychologists}}.
\newblock {\emph{\JournalTitle{British Journal of Clinical Psychology}}} \textbf{\bibinfo{volume}{61}}, \bibinfo{pages}{130--135} (\bibinfo{year}{2022}).

\bibitem{deangelis2023funding}
\bibinfo{author}{DeAngelis, T.}
\newblock \bibinfo{journal}{\bibinfo{title}{As funding cools, venture capitalists shift investments in mental health}}.
\newblock {\emph{\JournalTitle{American Psychological Association 2023 Trends Report}}} \textbf{\bibinfo{volume}{54}}, \bibinfo{pages}{83} (\bibinfo{year}{2023}).

\bibitem{kim2023databrokers}
\bibinfo{author}{Kim, J.}
\newblock \bibinfo{journal}{\bibinfo{title}{Data brokers and the sale of americans’ mental health data: The exchange of our most sensitive data and what it means for personal privacy}}.
\newblock {\emph{\JournalTitle{Duke Sanford Cyber Policy Program}}}  (\bibinfo{year}{2023}).

\bibitem{bossewitch2022digital}
\bibinfo{author}{Bossewitch, J.} \emph{et~al.}
\newblock \bibinfo{title}{Digital futures in mind: Reflecting on technological experiments in mental health \& crisis support}.
\newblock \bibinfo{howpublished}{\url{https://automatingmentalhealth.cc/}} (\bibinfo{year}{2022}).

\bibitem{nbcnews2023koko}
\bibinfo{author}{Ingram, D.}
\newblock \bibinfo{journal}{\bibinfo{title}{A mental health tech company ran an ai experiment on real users. nothing’s stopping apps from conducting more.}}
\newblock {\emph{\JournalTitle{NBC News}}}  (\bibinfo{year}{2023}).

\bibitem{politico2022ctl}
\bibinfo{author}{Hendel, J.}
\newblock \bibinfo{journal}{\bibinfo{title}{Crisis text line ends data-sharing relationship with for-profit spinoff}}.
\newblock {\emph{\JournalTitle{Politico}}}  (\bibinfo{year}{2022}).

\bibitem{warren2022senate}
\bibinfo{author}{{Office of U.S. Senator Elizabeth Warren}}.
\newblock \bibinfo{journal}{\bibinfo{title}{Warren, booker, wyden call on mental health apps to provide answers on data privacy and sharing practices that may put patients’ data at risk of exploitation}}.
\newblock {\emph{\JournalTitle{U.S. Senate}}}  (\bibinfo{year}{2022}).

\bibitem{ftcbetterhelp2023}
\bibinfo{author}{{Federal Trade Commission}}.
\newblock \bibinfo{journal}{\bibinfo{title}{Ftc to ban betterhelp from revealing consumers’ data, including sensitive mental health information, to facebook and others for targeted advertising}}.
\newblock {\emph{\JournalTitle{Press Release, Federal Trade Commission}}}  (\bibinfo{year}{2023}).

\bibitem{tiktokhearings2023}
\bibinfo{author}{Chan, K.}, \bibinfo{author}{Hadero, H.} \& \bibinfo{author}{Amiri, F.}
\newblock \bibinfo{title}{Tiktok ceo testifies before house committee over security concerns, connection to china}.
\newblock In \emph{\bibinfo{booktitle}{PBS NewsHour}} (\bibinfo{year}{2023}).

\bibitem{wykes2019towards}
\bibinfo{author}{Wykes, T.}, \bibinfo{author}{Lipshitz, J.} \& \bibinfo{author}{Schueller, S.~M.}
\newblock \bibinfo{journal}{\bibinfo{title}{Towards the design of ethical standards related to digital mental health and all its applications}}.
\newblock {\emph{\JournalTitle{Current Treatment Options in Psychiatry}}} \textbf{\bibinfo{volume}{6}}, \bibinfo{pages}{232--242} (\bibinfo{year}{2019}).

\bibitem{jilka2021terms}
\bibinfo{author}{Jilka, S.} \emph{et~al.}
\newblock \bibinfo{journal}{\bibinfo{title}{Terms and conditions apply: Critical issues for readability and jargon in mental health depression apps}}.
\newblock {\emph{\JournalTitle{Internet Interventions}}} \textbf{\bibinfo{volume}{25}}, \bibinfo{pages}{100433} (\bibinfo{year}{2021}).

\bibitem{parker2019private}
\bibinfo{author}{Parker, L.}, \bibinfo{author}{Halter, V.}, \bibinfo{author}{Karliychuk, T.} \& \bibinfo{author}{Grundy, Q.}
\newblock \bibinfo{journal}{\bibinfo{title}{How private is your mental health app data? an empirical study of mental health app privacy policies and practices}}.
\newblock {\emph{\JournalTitle{International journal of law and psychiatry}}} \textbf{\bibinfo{volume}{64}}, \bibinfo{pages}{198--204} (\bibinfo{year}{2019}).

\bibitem{obar2020biggest}
\bibinfo{author}{Obar, J.~A.} \& \bibinfo{author}{Oeldorf-Hirsch, A.}
\newblock \bibinfo{journal}{\bibinfo{title}{The biggest lie on the internet: Ignoring the privacy policies and terms of service policies of social networking services}}.
\newblock {\emph{\JournalTitle{Information, Communication \& Society}}} \textbf{\bibinfo{volume}{23}}, \bibinfo{pages}{128--147} (\bibinfo{year}{2020}).

\bibitem{ienca2022mental}
\bibinfo{author}{Ienca, M.} \& \bibinfo{author}{Malgieri, G.}
\newblock \bibinfo{journal}{\bibinfo{title}{Mental data protection and the gdpr}}.
\newblock {\emph{\JournalTitle{Journal of Law and the Biosciences}}} \textbf{\bibinfo{volume}{9}}, \bibinfo{pages}{lsac006} (\bibinfo{year}{2022}).

\bibitem{gooding2019mapping}
\bibinfo{author}{Gooding, P.}
\newblock \bibinfo{journal}{\bibinfo{title}{Mapping the rise of digital mental health technologies: Emerging issues for law and society}}.
\newblock {\emph{\JournalTitle{International journal of law and psychiatry}}} \textbf{\bibinfo{volume}{67}}, \bibinfo{pages}{101498} (\bibinfo{year}{2019}).

\bibitem{im2021yes}
\bibinfo{author}{Im, J.} \emph{et~al.}
\newblock \bibinfo{title}{Yes: Affirmative consent as a theoretical framework for understanding and imagining social platforms}.
\newblock In \emph{\bibinfo{booktitle}{Proceedings of the 2021 CHI Conference on Human Factors in Computing Systems}}, \bibinfo{pages}{1--18} (\bibinfo{year}{2021}).

\bibitem{friedman2019yes}
\bibinfo{author}{Friedman, J.} \& \bibinfo{author}{Valenti, J.}
\newblock \emph{\bibinfo{title}{Yes means yes!: Visions of female sexual power and a world without rape}} (\bibinfo{publisher}{Seal Press}, \bibinfo{year}{2019}).

\bibitem{hilgert2016burden}
\bibinfo{author}{Hilgert, N.}
\newblock \bibinfo{journal}{\bibinfo{title}{The burden of consent: Due process and the emerging adoption of the affirmative consent standard in sexual assault laws}}.
\newblock {\emph{\JournalTitle{Ariz. L. Rev.}}} \textbf{\bibinfo{volume}{58}}, \bibinfo{pages}{867} (\bibinfo{year}{2016}).

\bibitem{saks2010refusing}
\bibinfo{author}{Saks, E.~R.}
\newblock \emph{\bibinfo{title}{Refusing care: Forced treatment and the rights of the mentally ill}} (\bibinfo{publisher}{University of Chicago Press}, \bibinfo{year}{2010}).

\bibitem{saks2007center}
\bibinfo{author}{Saks, E.~R.}
\newblock \emph{\bibinfo{title}{The center cannot hold: My journey through madness}} (\bibinfo{publisher}{Hachette UK}, \bibinfo{year}{2007}).

\bibitem{wipond2023your}
\bibinfo{author}{Wipond, R.}
\newblock \emph{\bibinfo{title}{Your Consent is Not Required: The Rise in Psychiatric Detentions, Forced Treatment, and Abusive Guardianships}} (\bibinfo{publisher}{BenBella Books}, \bibinfo{year}{2023}).

\bibitem{porter1990foucault}
\bibinfo{author}{Porter, R.}
\newblock \bibinfo{journal}{\bibinfo{title}{Foucault's great confinement}}.
\newblock {\emph{\JournalTitle{History of the Human Sciences}}} \textbf{\bibinfo{volume}{3}}, \bibinfo{pages}{47--54} (\bibinfo{year}{1990}).

\bibitem{andrews1998case}
\bibinfo{author}{Andrews, J.}
\newblock \bibinfo{journal}{\bibinfo{title}{Case notes, case histories, and the patient's experience of insanity at gartnavel royal asylum, glasgow, in the nineteenth century}}.
\newblock {\emph{\JournalTitle{Social History of Medicine}}} \textbf{\bibinfo{volume}{11}}, \bibinfo{pages}{255--281} (\bibinfo{year}{1998}).

\bibitem{reid1989history}
\bibinfo{author}{Reid-Green, K.~S.}
\newblock \bibinfo{journal}{\bibinfo{title}{The history of census tabulation}}.
\newblock {\emph{\JournalTitle{Scientific American}}} \textbf{\bibinfo{volume}{260}}, \bibinfo{pages}{98--103} (\bibinfo{year}{1989}).

\bibitem{hanganu2019diagnosing}
\bibinfo{author}{Hanganu-Bresch, C.} \& \bibinfo{author}{Berkenkotter, C.}
\newblock \emph{\bibinfo{title}{Diagnosing madness: The discursive construction of the psychiatric patient, 1850-1920}} (\bibinfo{publisher}{Univ of South Carolina Press}, \bibinfo{year}{2019}).

\bibitem{wright2022asylums}
\bibinfo{author}{Wright, D.}
\newblock \bibinfo{title}{Asylums and alienists: The institutional foundations of psychiatry, 1760--1914}.
\newblock In \emph{\bibinfo{booktitle}{The Palgrave Handbook of the History of Human Sciences}}, \bibinfo{pages}{1--20} (\bibinfo{publisher}{Springer}, \bibinfo{year}{2022}).

\bibitem{forret2016deaf}
\bibinfo{author}{Forret, J.}
\newblock \bibinfo{journal}{\bibinfo{title}{" deaf \& dumb, blind, insane, or idiotic": The census, slaves, and disability in the late antebellum south}}.
\newblock {\emph{\JournalTitle{The Journal of Southern History}}} \textbf{\bibinfo{volume}{82}}, \bibinfo{pages}{503--548} (\bibinfo{year}{2016}).

\bibitem{horwitz2011checkered}
\bibinfo{author}{Horwitz, A.~V.} \& \bibinfo{author}{Grob, G.~N.}
\newblock \bibinfo{journal}{\bibinfo{title}{The checkered history of american psychiatric epidemiology}}.
\newblock {\emph{\JournalTitle{The Milbank Quarterly}}} \textbf{\bibinfo{volume}{89}}, \bibinfo{pages}{628--657} (\bibinfo{year}{2011}).

\bibitem{deutsch1944first}
\bibinfo{author}{Deutsch, A.}
\newblock \bibinfo{journal}{\bibinfo{title}{The first us census of the insane (1840) and its use as pro-slavery propaganda}}.
\newblock {\emph{\JournalTitle{Bulletin of the History of Medicine}}} \textbf{\bibinfo{volume}{15}}, \bibinfo{pages}{469--482} (\bibinfo{year}{1944}).

\bibitem{freud1997dora}
\bibinfo{author}{Freud, S.}
\newblock \emph{\bibinfo{title}{Dora: An analysis of a case of hysteria}} (\bibinfo{publisher}{Simon and Schuster}, \bibinfo{year}{1997}).

\bibitem{lipton1991analyst}
\bibinfo{author}{Lipton, E.~L.}
\newblock \bibinfo{journal}{\bibinfo{title}{The analyst's use of clinical data, and other issues of confidentiality}}.
\newblock {\emph{\JournalTitle{Journal of the American Psychoanalytic Association}}} \textbf{\bibinfo{volume}{39}}, \bibinfo{pages}{967--985} (\bibinfo{year}{1991}).

\bibitem{freud1962fragment}
\bibinfo{author}{Freud, S.} \& \bibinfo{author}{Strachey, A.}
\newblock \emph{\bibinfo{title}{Fragment of an analysis of a case of hysteria (1905 [1901])}}, vol.~\bibinfo{volume}{7} (\bibinfo{publisher}{Hogarth Press}, \bibinfo{year}{1962}).

\bibitem{mosher2002ethical}
\bibinfo{author}{Mosher, P.~W.} \& \bibinfo{author}{Swire, P.~P.}
\newblock \bibinfo{journal}{\bibinfo{title}{The ethical and legal implications of jaffee v redmond and the hipaa medical privacy rule for psychotherapy and general psychiatry}}.
\newblock {\emph{\JournalTitle{Psychiatric Clinics}}} \textbf{\bibinfo{volume}{25}}, \bibinfo{pages}{A575--A584} (\bibinfo{year}{2002}).

\bibitem{leach1997psychological}
\bibinfo{author}{Leach, M.~M.} \& \bibinfo{author}{Harbin, J.~J.}
\newblock \bibinfo{journal}{\bibinfo{title}{Psychological ethics codes: A comparison of twenty-four countries}}.
\newblock {\emph{\JournalTitle{International Journal of Psychology}}} \textbf{\bibinfo{volume}{32}}, \bibinfo{pages}{181--192} (\bibinfo{year}{1997}).

\bibitem{bondre2021protecting}
\bibinfo{author}{Bondre, A.}, \bibinfo{author}{Pathare, S.} \& \bibinfo{author}{Naslund, J.~A.}
\newblock \bibinfo{journal}{\bibinfo{title}{Protecting mental health data privacy in india: The case of data linkage with aadhaar}}.
\newblock {\emph{\JournalTitle{Global Health: Science and Practice}}} \textbf{\bibinfo{volume}{9}}, \bibinfo{pages}{467--480} (\bibinfo{year}{2021}).

\bibitem{lally2019mental}
\bibinfo{author}{Lally, J.}, \bibinfo{author}{Samaniego, R.~M.} \& \bibinfo{author}{Tully, J.}
\newblock \bibinfo{journal}{\bibinfo{title}{Mental health legislation in the philippines: Philippine mental health act}}.
\newblock {\emph{\JournalTitle{BJPsych international}}} \textbf{\bibinfo{volume}{16}}, \bibinfo{pages}{65--67} (\bibinfo{year}{2019}).

\bibitem{GDPR2016a}
\bibinfo{author}{{European Parliament}} \& \bibinfo{author}{{Council of the European Union}}.
\newblock \bibinfo{title}{Regulation ({EU}) 2016/679 of the {European} {Parliament} and of the {Council}}.

\bibitem{borecky2019reweighing}
\bibinfo{author}{Borecky, A.}, \bibinfo{author}{Thomsen, C.} \& \bibinfo{author}{Dubov, A.}
\newblock \bibinfo{journal}{\bibinfo{title}{Reweighing the ethical tradeoffs in the involuntary hospitalization of suicidal patients}}.
\newblock {\emph{\JournalTitle{The American Journal of Bioethics}}} \textbf{\bibinfo{volume}{19}}, \bibinfo{pages}{71--83} (\bibinfo{year}{2019}).

\bibitem{beauchamp2001principles}
\bibinfo{author}{Beauchamp, T.~L.} \& \bibinfo{author}{Childress, J.~F.}
\newblock \emph{\bibinfo{title}{Principles of biomedical ethics}} (\bibinfo{publisher}{Oxford University Press, USA}, \bibinfo{year}{2001}).

\bibitem{united1978belmont}
\bibinfo{author}{for the Protection of Human Subjects~of Biomedical, U. S. N.~C.} \& \bibinfo{author}{Research, B.}
\newblock \emph{\bibinfo{title}{The Belmont report: ethical principles and guidelines for the protection of human subjects of research}}, vol.~\bibinfo{volume}{1} (\bibinfo{publisher}{Department of Health, Education, and Welfare, National Commission for the~…}, \bibinfo{year}{1978}).

\bibitem{spellecy2021history}
\bibinfo{author}{Spellecy, R.} \& \bibinfo{author}{Busse, K.}
\newblock \bibinfo{journal}{\bibinfo{title}{The history of human subjects research and rationale for institutional review board oversight}}.
\newblock {\emph{\JournalTitle{Nutrition in Clinical Practice}}} \textbf{\bibinfo{volume}{36}}, \bibinfo{pages}{560--567} (\bibinfo{year}{2021}).

\bibitem{zeavin2021distance}
\bibinfo{author}{Zeavin, H.}
\newblock \emph{\bibinfo{title}{The distance cure: a history of teletherapy}} (\bibinfo{publisher}{MIT Press}, \bibinfo{year}{2021}).

\bibitem{zeavinthirdchoice2020}
\bibinfo{author}{Zeavin, H.}
\newblock \bibinfo{journal}{\bibinfo{title}{The third choice: Suicide hotlines, psychiatry, and the police}}.
\newblock {\emph{\JournalTitle{Somatosphere}}}  (\bibinfo{year}{2020}).

\bibitem{aas2006}
\bibinfo{author}{{American Association of Suicidology}}.
\newblock \bibinfo{title}{Organization accreditation standards manual} (\bibinfo{year}{2006}).

\bibitem{nsp2010}
\bibinfo{author}{Lifeline, N. S.~P.}
\newblock \bibinfo{title}{National suicide prevention lifeline: Policy for helping callers at imminent risk} (\bibinfo{year}{2010}).

\bibitem{testa2010civil}
\bibinfo{author}{Testa, M.} \& \bibinfo{author}{West, S.~G.}
\newblock \bibinfo{journal}{\bibinfo{title}{Civil commitment in the united states}}.
\newblock {\emph{\JournalTitle{Psychiatry (Edgmont)}}} \textbf{\bibinfo{volume}{7}}, \bibinfo{pages}{30} (\bibinfo{year}{2010}).

\bibitem{naghavi2018global}
\bibinfo{author}{Naghavi, M.} \emph{et~al.}
\newblock \bibinfo{journal}{\bibinfo{title}{Global mortality from firearms, 1990-2016}}.
\newblock {\emph{\JournalTitle{Jama}}} \textbf{\bibinfo{volume}{320}}, \bibinfo{pages}{792--814} (\bibinfo{year}{2018}).

\bibitem{rivara2018firearm}
\bibinfo{author}{Rivara, F.~P.}, \bibinfo{author}{Studdert, D.~M.} \& \bibinfo{author}{Wintemute, G.~J.}
\newblock \bibinfo{journal}{\bibinfo{title}{Firearm-related mortality: a global public health problem}}.
\newblock {\emph{\JournalTitle{Jama}}} \textbf{\bibinfo{volume}{320}}, \bibinfo{pages}{764--765} (\bibinfo{year}{2018}).

\bibitem{ilic2022worldwide}
\bibinfo{author}{Ilic, I.}, \bibinfo{author}{Zivanovic~Macuzic, I.}, \bibinfo{author}{Kocic, S.} \& \bibinfo{author}{Ilic, M.}
\newblock \bibinfo{journal}{\bibinfo{title}{Worldwide suicide mortality trends by firearm (1990--2019): A joinpoint regression analysis}}.
\newblock {\emph{\JournalTitle{PLoS one}}} \textbf{\bibinfo{volume}{17}}, \bibinfo{pages}{e0267817} (\bibinfo{year}{2022}).

\bibitem{teplin2000keeping}
\bibinfo{author}{Teplin, L.~A.}
\newblock \bibinfo{journal}{\bibinfo{title}{Keeping the peace: Police discretion and mentally ill persons.}}
\newblock {\emph{\JournalTitle{National institute of justice journal}}} \textbf{\bibinfo{volume}{244}}, \bibinfo{pages}{8--15} (\bibinfo{year}{2000}).

\bibitem{leppert2022suicide}
\bibinfo{author}{Leppert, M.} \& \bibinfo{author}{Jervert, K.}
\newblock \bibinfo{journal}{\bibinfo{title}{Suicide hotlines and the impact of non-consensual interventions}}.
\newblock {\emph{\JournalTitle{Mad in America}}}  (\bibinfo{year}{2022}).

\bibitem{FTCComplaint2023}
\bibinfo{author}{{Federal Trade Commission}}.
\newblock \bibinfo{title}{{FTC v. BetterHelp Inc.}}
\newblock \bibinfo{howpublished}{FTC Complaint} (\bibinfo{year}{2023}).

\bibitem{powell2018complexity}
\bibinfo{author}{Powell, A.}, \bibinfo{author}{Singh, P.}, \bibinfo{author}{Torous, J.} \emph{et~al.}
\newblock \bibinfo{journal}{\bibinfo{title}{The complexity of mental health app privacy policies: a potential barrier to privacy}}.
\newblock {\emph{\JournalTitle{JMIR mHealth and uHealth}}} \textbf{\bibinfo{volume}{6}}, \bibinfo{pages}{e9871} (\bibinfo{year}{2018}).

\bibitem{iwaya2023privacy}
\bibinfo{author}{Iwaya, L.~H.}, \bibinfo{author}{Babar, M.~A.}, \bibinfo{author}{Rashid, A.} \& \bibinfo{author}{Wijayarathna, C.}
\newblock \bibinfo{journal}{\bibinfo{title}{On the privacy of mental health apps: An empirical investigation and its implications for app development}}.
\newblock {\emph{\JournalTitle{Empirical Software Engineering}}} \textbf{\bibinfo{volume}{28}}, \bibinfo{pages}{2} (\bibinfo{year}{2023}).

\bibitem{kendall2023nhs}
\bibinfo{author}{Kendall, T.}
\newblock \bibinfo{title}{Nhs england position on serenity integrated mentoring (sim) and similar models}.
\newblock \bibinfo{howpublished}{\url{https://www.england.nhs.uk/long-read/nhs-england-position-on-serenity-integrated-mentoring-and-similar-models/}} (\bibinfo{year}{2023}).
\newblock \bibinfo{note}{Publication reference: PRN00317}.

\bibitem{middleton2014systematic}
\bibinfo{author}{Middleton, A.}, \bibinfo{author}{Gunn, J.}, \bibinfo{author}{Bassilios, B.} \& \bibinfo{author}{Pirkis, J.}
\newblock \bibinfo{journal}{\bibinfo{title}{Systematic review of research into frequent callers to crisis helplines}}.
\newblock {\emph{\JournalTitle{Journal of Telemedicine and Telecare}}} \textbf{\bibinfo{volume}{20}}, \bibinfo{pages}{89--98} (\bibinfo{year}{2014}).

\bibitem{wash_state_law_1155}
\bibinfo{title}{Engrossed substitute house bill 1155: Consumer health data} (\bibinfo{year}{2023}).
\newblock \bibinfo{note}{Effective date: July 23, 2023}.

\bibitem{FTC_2023}
\bibinfo{author}{{Federal Trade Commission}}.
\newblock \bibinfo{title}{Ftc proposes amendments to strengthen and modernize the health breach notification rule}.
\newblock \bibinfo{howpublished}{\url{https://www.ftc.gov/news-events/news/press-releases/2023/05/ftc-proposes-amendments-strengthen-modernize-health-breach-notification-rule}} (\bibinfo{year}{2023}).

\bibitem{mubarak2022promoting}
\bibinfo{author}{Mubarak, E.}, \bibinfo{author}{Turner, V.}, \bibinfo{author}{Shuman, A.~G.}, \bibinfo{author}{Firn, J.} \& \bibinfo{author}{Price, D.}
\newblock \bibinfo{journal}{\bibinfo{title}{Promoting antiracist mental health crisis responses}}.
\newblock {\emph{\JournalTitle{AMA Journal of Ethics}}} \textbf{\bibinfo{volume}{24}}, \bibinfo{pages}{788--794} (\bibinfo{year}{2022}).

\bibitem{rochford2023leveraging}
\bibinfo{author}{Rochford, B.}, \bibinfo{author}{Pendse, S.}, \bibinfo{author}{Kumar, N.} \& \bibinfo{author}{De~Choudhury, M.}
\newblock \bibinfo{journal}{\bibinfo{title}{Leveraging symptom search data to understand disparities in us mental health care: demographic analysis of search engine trace data}}.
\newblock {\emph{\JournalTitle{JMIR Mental Health}}} \textbf{\bibinfo{volume}{10}}, \bibinfo{pages}{e43253} (\bibinfo{year}{2023}).

\bibitem{borghouts2021barriers}
\bibinfo{author}{Borghouts, J.} \emph{et~al.}
\newblock \bibinfo{journal}{\bibinfo{title}{Barriers to and facilitators of user engagement with digital mental health interventions: systematic review}}.
\newblock {\emph{\JournalTitle{Journal of medical Internet research}}} \textbf{\bibinfo{volume}{23}}, \bibinfo{pages}{e24387} (\bibinfo{year}{2021}).

\bibitem{sanchez2019can}
\bibinfo{author}{Sanchez-Rola, I.} \emph{et~al.}
\newblock \bibinfo{title}{Can i opt out yet? gdpr and the global illusion of cookie control}.
\newblock In \emph{\bibinfo{booktitle}{Proceedings of the 2019 ACM Asia conference on computer and communications security}}, \bibinfo{pages}{340--351} (\bibinfo{year}{2019}).

\bibitem{NHS_Digital_2023}
\bibinfo{author}{{NHS Digital}}.
\newblock \bibinfo{title}{Understanding the national data opt-out}.
\newblock \bibinfo{howpublished}{\url{https://digital.nhs.uk/services/national-data-opt-out/understanding-the-national-data-opt-out}} (\bibinfo{year}{2023}).

\bibitem{mackay2015opt}
\bibinfo{author}{MacKay, D.}
\newblock \bibinfo{journal}{\bibinfo{title}{Opt-out and consent}}.
\newblock {\emph{\JournalTitle{Journal of Medical Ethics}}}  (\bibinfo{year}{2015}).

\bibitem{988lifeline_lived_experience}
\bibinfo{author}{{988 Suicide \& Crisis Lifeline}}.
\newblock \bibinfo{title}{Lived experience committee}.
\newblock \bibinfo{howpublished}{\url{https://988lifeline.org/lived-experience-committee/}} (\bibinfo{year}{2024}).

\bibitem{samhsa_advisory_councils}
\bibinfo{author}{{Substance Abuse and Mental Health Services Administration (SAMHSA)}}.
\newblock \bibinfo{title}{Advisory councils}.
\newblock \bibinfo{howpublished}{\url{https://www.samhsa.gov/about-us/advisory-councils}} (\bibinfo{year}{2024}).

\bibitem{george2014systematic}
\bibinfo{author}{George, S.}, \bibinfo{author}{Duran, N.} \& \bibinfo{author}{Norris, K.}
\newblock \bibinfo{journal}{\bibinfo{title}{A systematic review of barriers and facilitators to minority research participation among african americans, latinos, asian americans, and pacific islanders}}.
\newblock {\emph{\JournalTitle{American journal of public health}}} \textbf{\bibinfo{volume}{104}}, \bibinfo{pages}{e16--e31} (\bibinfo{year}{2014}).

\bibitem{mckim2023meaningful}
\bibinfo{author}{McKim, C.}
\newblock \bibinfo{journal}{\bibinfo{title}{Meaningful member-checking: A structured approach to member-checking}}.
\newblock {\emph{\JournalTitle{American Journal of Qualitative Research}}} \textbf{\bibinfo{volume}{7}}, \bibinfo{pages}{41--52} (\bibinfo{year}{2023}).

\bibitem{lincoln1985naturalistic}
\bibinfo{author}{Lincoln, Y.~S.} \& \bibinfo{author}{Guba, E.~G.}
\newblock \emph{\bibinfo{title}{Naturalistic inquiry}} (\bibinfo{publisher}{Sage}, \bibinfo{year}{1985}).

\bibitem{sastre2021decision}
\bibinfo{author}{Sastre-Buades, A.}, \bibinfo{author}{Alacreu-Crespo, A.}, \bibinfo{author}{Courtet, P.}, \bibinfo{author}{Baca-Garcia, E.} \& \bibinfo{author}{Barrigon, M.~L.}
\newblock \bibinfo{journal}{\bibinfo{title}{Decision-making in suicidal behavior: A systematic review and meta-analysis}}.
\newblock {\emph{\JournalTitle{Neuroscience \& Biobehavioral Reviews}}} \textbf{\bibinfo{volume}{131}}, \bibinfo{pages}{642--662} (\bibinfo{year}{2021}).

\bibitem{sandler2020mobile}
\bibinfo{author}{Sandler, R.~D.}
\newblock \bibinfo{journal}{\bibinfo{title}{Mobile medical id: A resource for the off-duty clinician}}.
\newblock {\emph{\JournalTitle{The Journal of Emergency Medicine}}} \textbf{\bibinfo{volume}{59}}, \bibinfo{pages}{141--142} (\bibinfo{year}{2020}).

\bibitem{gupta2021securing}
\bibinfo{author}{Gupta, M.}, \bibinfo{author}{Jain, R.}, \bibinfo{author}{Kumari, M.} \& \bibinfo{author}{Narula, G.}
\newblock \bibinfo{journal}{\bibinfo{title}{Securing healthcare data by using blockchain}}.
\newblock {\emph{\JournalTitle{Applications of blockchain in healthcare}}} \bibinfo{pages}{93--114} (\bibinfo{year}{2021}).

\bibitem{yue2016healthcare}
\bibinfo{author}{Yue, X.}, \bibinfo{author}{Wang, H.}, \bibinfo{author}{Jin, D.}, \bibinfo{author}{Li, M.} \& \bibinfo{author}{Jiang, W.}
\newblock \bibinfo{journal}{\bibinfo{title}{Healthcare data gateways: found healthcare intelligence on blockchain with novel privacy risk control}}.
\newblock {\emph{\JournalTitle{Journal of medical systems}}} \textbf{\bibinfo{volume}{40}}, \bibinfo{pages}{1--8} (\bibinfo{year}{2016}).

\bibitem{sarkar2018towards}
\bibinfo{author}{Sarkar, S.}, \bibinfo{author}{Banatre, J.-P.}, \bibinfo{author}{Rilling, L.} \& \bibinfo{author}{Morin, C.}
\newblock \bibinfo{title}{Towards enforcement of the eu gdpr: enabling data erasure}.
\newblock In \emph{\bibinfo{booktitle}{2018 IEEE International Conference on Internet of Things (iThings) and IEEE Green Computing and Communications (GreenCom) and IEEE Cyber, Physical and Social Computing (CPSCom) and IEEE Smart Data (SmartData)}}, \bibinfo{pages}{222--229} (\bibinfo{organization}{IEEE}, \bibinfo{year}{2018}).

\bibitem{simon2019assessing}
\bibinfo{author}{Simon, G.~E.} \emph{et~al.}
\newblock \bibinfo{journal}{\bibinfo{title}{Assessing and minimizing re-identification risk in research data derived from health care records}}.
\newblock {\emph{\JournalTitle{eGEMs}}} \textbf{\bibinfo{volume}{7}} (\bibinfo{year}{2019}).

\bibitem{niinimaki2019representation}
\bibinfo{author}{Niinim{\"a}ki, T.}, \bibinfo{author}{Heikkil{\"a}, M.~A.}, \bibinfo{author}{Honkela, A.} \& \bibinfo{author}{Kaski, S.}
\newblock \bibinfo{journal}{\bibinfo{title}{Representation transfer for differentially private drug sensitivity prediction}}.
\newblock {\emph{\JournalTitle{Bioinformatics}}} \textbf{\bibinfo{volume}{35}}, \bibinfo{pages}{i218--i224} (\bibinfo{year}{2019}).

\bibitem{liu2020blockchain}
\bibinfo{author}{Liu, X.}, \bibinfo{author}{Zhou, P.}, \bibinfo{author}{Qiu, T.} \& \bibinfo{author}{Wu, D.~O.}
\newblock \bibinfo{journal}{\bibinfo{title}{Blockchain-enabled contextual online learning under local differential privacy for coronary heart disease diagnosis in mobile edge computing}}.
\newblock {\emph{\JournalTitle{IEEE Journal of Biomedical and Health Informatics}}} \textbf{\bibinfo{volume}{24}}, \bibinfo{pages}{2177--2188} (\bibinfo{year}{2020}).

\bibitem{lv2021security}
\bibinfo{author}{Lv, Z.} \& \bibinfo{author}{Piccialli, F.}
\newblock \bibinfo{journal}{\bibinfo{title}{The security of medical data on internet based on differential privacy technology}}.
\newblock {\emph{\JournalTitle{ACM Transactions on Internet Technology}}} \textbf{\bibinfo{volume}{21}}, \bibinfo{pages}{1--18} (\bibinfo{year}{2021}).

\bibitem{roy2022interpretability}
\bibinfo{author}{Roy, T.~S.}, \bibinfo{author}{Basu, P.}, \bibinfo{author}{Priyanshu, A.} \& \bibinfo{author}{Naidu, R.}
\newblock \bibinfo{journal}{\bibinfo{title}{Interpretability of fine-grained classification of sadness and depression}}.
\newblock {\emph{\JournalTitle{arXiv preprint arXiv:2203.10432}}}  (\bibinfo{year}{2022}).

\bibitem{basu2021benchmarking}
\bibinfo{author}{Basu, P.} \emph{et~al.}
\newblock \bibinfo{journal}{\bibinfo{title}{Benchmarking differential privacy and federated learning for bert models}}.
\newblock {\emph{\JournalTitle{arXiv preprint arXiv:2106.13973}}}  (\bibinfo{year}{2021}).

\bibitem{antunes2022federated}
\bibinfo{author}{Antunes, R.~S.}, \bibinfo{author}{Andr{\'e}~da Costa, C.}, \bibinfo{author}{K{\"u}derle, A.}, \bibinfo{author}{Yari, I.~A.} \& \bibinfo{author}{Eskofier, B.}
\newblock \bibinfo{journal}{\bibinfo{title}{Federated learning for healthcare: Systematic review and architecture proposal}}.
\newblock {\emph{\JournalTitle{ACM Transactions on Intelligent Systems and Technology (TIST)}}} \textbf{\bibinfo{volume}{13}}, \bibinfo{pages}{1--23} (\bibinfo{year}{2022}).

\bibitem{brauneck2023federated}
\bibinfo{author}{Brauneck, A.} \emph{et~al.}
\newblock \bibinfo{journal}{\bibinfo{title}{Federated machine learning in data-protection-compliant research}}.
\newblock {\emph{\JournalTitle{Nature Machine Intelligence}}} \textbf{\bibinfo{volume}{5}}, \bibinfo{pages}{2--4} (\bibinfo{year}{2023}).

\bibitem{dechand2019encryption}
\bibinfo{author}{Dechand, S.}, \bibinfo{author}{Naiakshina, A.}, \bibinfo{author}{Danilova, A.} \& \bibinfo{author}{Smith, M.}
\newblock \bibinfo{title}{In encryption we don’t trust: The effect of end-to-end encryption to the masses on user perception}.
\newblock In \emph{\bibinfo{booktitle}{2019 IEEE European Symposium on Security and Privacy (EuroS\&P)}}, \bibinfo{pages}{401--415} (\bibinfo{organization}{IEEE}, \bibinfo{year}{2019}).

\bibitem{kalkman2022patients}
\bibinfo{author}{Kalkman, S.} \emph{et~al.}
\newblock \bibinfo{journal}{\bibinfo{title}{Patients’ and public views and attitudes towards the sharing of health data for research: a narrative review of the empirical evidence}}.
\newblock {\emph{\JournalTitle{Journal of medical ethics}}} \textbf{\bibinfo{volume}{48}}, \bibinfo{pages}{3--13} (\bibinfo{year}{2022}).

\bibitem{sieberts2023young}
\bibinfo{author}{Sieberts, S.~K.} \emph{et~al.}
\newblock \bibinfo{journal}{\bibinfo{title}{Young people’s data governance preferences for their mental health data: Mindkind study findings from india, south africa, and the united kingdom}}.
\newblock {\emph{\JournalTitle{PloS one}}} \textbf{\bibinfo{volume}{18}}, \bibinfo{pages}{e0279857} (\bibinfo{year}{2023}).

\bibitem{olavsrud2021data}
\bibinfo{author}{Olavsrud, T.}
\newblock \bibinfo{journal}{\bibinfo{title}{Data governance: A best practices framework for managing data assets}}.
\newblock {\emph{\JournalTitle{CIO}}}  (\bibinfo{year}{2021}).

\bibitem{ada2021participatory}
\bibinfo{author}{{Ada Lovelace Institute}}.
\newblock \bibinfo{title}{Participatory data stewardship: A framework for involving people in the use of data} (\bibinfo{year}{2021}).

\bibitem{campbell2005historical}
\bibinfo{author}{Campbell, J.}
\newblock \bibinfo{journal}{\bibinfo{title}{The historical and philosophical development of peer-run support programs}}.
\newblock {\emph{\JournalTitle{On our own, together: Peer programs for people with mental illness}}} \textbf{\bibinfo{volume}{17}}, \bibinfo{pages}{66} (\bibinfo{year}{2005}).

\bibitem{thrul2022web3}
\bibinfo{author}{Thrul, J.}, \bibinfo{author}{Kalb, L.~G.}, \bibinfo{author}{Finan, P.~H.}, \bibinfo{author}{Prager, Z.} \& \bibinfo{author}{Naslund, J.~A.}
\newblock \bibinfo{journal}{\bibinfo{title}{Web3 and digital mental health: Opportunities to scale sustainable mental health promotion and peer support}}.
\newblock {\emph{\JournalTitle{Frontiers in Psychiatry}}} \bibinfo{pages}{1608} (\bibinfo{year}{2022}).

\bibitem{temanararaunga}
\bibinfo{author}{Network, T. M. R. M. D.~S.}
\newblock \bibinfo{title}{Principles of m{\=a}ori data sovereignty} (\bibinfo{year}{2018}).

\bibitem{feygin2021data}
\bibinfo{author}{Feygin, Y.} \emph{et~al.}
\newblock \bibinfo{journal}{\bibinfo{title}{A data dividend that works: steps toward building an equitable data economy}}.
\newblock {\emph{\JournalTitle{Berggruen Institute}}}  (\bibinfo{year}{2021}).

\end{thebibliography}
\end{document}